\documentclass[superscriptaddress,aps,pre,twocolumn,showpacs,preprintnumbers,longbibliography]{revtex4-1}
\usepackage{amsmath}
\usepackage{times,mathptmx,amssymb}
\usepackage{graphicx}
\usepackage{xcolor}

\usepackage{float}

\newcommand{\derpar}[2]{\frac{\partial #1}{\partial #2}}
\newcommand{\VEC}[1]{\mathbf{#1}}
\newcommand{\HAT}[1]{\hat{\mathbf{#1}}}

\begin{document}
\title{Vertex model instabilities for  tissues subject to cellular activity or applied stresses}

\author{Fernanda P\'erez-Verdugo}
\affiliation{Departamento de F\'{\i}sica, FCFM, Universidad de Chile, Santiago, Chile}

\author{Jean--Francois Joanny}
\affiliation{Coll\`ege de France, 11 place Marcelin Berthelot, 75005 Paris, France }
\affiliation{Institut Curie PSL University 26 rue d'Ulm 75248 Paris Cedex 05}

\author{Rodrigo Soto}
\affiliation{Departamento de F\'{\i}sica, FCFM, Universidad de Chile, Santiago, Chile}

\begin{abstract}
The vertex model is widely used to describe the dynamics of epithelial tissues, because of its simplicity and versatility and the direct inclusion of biophysical parameters. Here, it is shown that quite generally, when cells modify their equilibrium perimeter due to their activity, or the tissue is subject to external stresses, the tissue becomes unstable with deformations that couple  pure-shear or deviatoric modes, with rotation and expansion modes. For short times, these instabilities deform cells increasing their ellipticity while, at longer times, cells become non-convex, indicating that the vertex model ceases to be a valid description for tissues under these conditions. The agreement between the analytic calculations  performed for a regular hexagonal tissue and the simulations of disordered tissues is excellent due to the homogenization of the tissue at long wavelengths.
\end{abstract}
\maketitle

\section{Introduction}
The vertex model, initially proposed to describe foams and soap bubbles 
\cite{weaire1984soap,okuzono1995intermittent}, has\ been extended to 
describe epithelial tissues \cite{nagai1988vertex,nagai2001dynamic,staple2010mechanics, fletcher2014vertex} with large 
success. Applications include the study of cell division \cite{mao2011planar}, tissue 
elongation \cite{rauzi2008nature} and epithelial packing in wing disk and ventral furrow 
formation in \textit{Drosophila} 
\cite{leptin1990cell,farhadifar2007influence,spahn2013vertex}, tube formation 
\cite{lubarsky2003tube,inoue2016mechanical}, and the rigidity transition in active 
tissues~\cite{bi2015density}. Approximating each $c$ cell as a polygon, an energy functional is 
built that penalizes the deviations of the actual cell areas $\left(A_c\right)$ and perimeters $\left(P_c\right)$ from preferred values ($A_{0c}$ and $P_{0c}$, respectively). 
In the most generic form, the  energy functional is  
\begin{align}
E =  \frac{K_A}{2}\sum_c\left( A_c - A_{0c}\right)^2 + \frac{K_P}{2}\sum_c\left( P_c -
P_{0c}\right)^2+J\sum_{\langle i,j \rangle} l_{ij},
\label{eq.Evertex}
\end{align}
with  $l_{ij}$ 
the length of the cell edge shared by vertices $i$ and $j$.
$K_A$ is the area elastic modulus, which describes the three dimensional 
incompressibility of the layer and the resistance to height fluctuations; $K_P$ is the 
perimeter elastic modulus related to the actin-myosin contractility; $J$ is the adhesion 
energy per unit length and represents a constant line tension. Although it is possible to absorb the last term into the 
second one by redefining $P_{0c}$, we opt to keep all terms, 
such that the different constants retain a direct interpretation.
Through this work we consider $A_{0c}$ and $P_{0c}$ given by the initial geometry of each cell. Hence, the model only has three free parameters.

In its usual form, the degrees of freedom of the model are the positions of the 
vertices  $\VEC r_i$, which evolve variationally as
\begin{align}
\frac{d \VEC r_i}{dt} = -\gamma \derpar{E}{\VEC r_i},
\label{eq.variationaldyn}
\end{align}
where $\gamma$ is a mobility that we will absorb in $K_A, K_P$ and $J$, which now have units of relaxation rates times different powers of length. 
 
 Active stresses are continuously induced  by cell divisions, 
 extrusions and rearrangements between neighboring cells \cite{etournay2015interplay}. 
 Also, stresses are generated by cell growth \cite{vincent2013cell} and 
 contractions \cite{han2018cell}; processes that can  be easily included in 
 the vertex model as changes in the equilibrium cell parameters. 
 
 In Refs.~\cite{farhadifar2007influence, staple2010mechanics}, the vertex model 
 was used to obtain the phase diagram of the ground state (the most 
 relaxed network configuration) of a proliferating tissue, initially made of a 
 regular hexagonal packing. They found a phase 
 transition induced by cell division in the parameter space $[J/(K_A A_{0c}^{3/2}), 
K_P/(K_AA_{0c})]$. One phase corresponds to a single ground state, with 
 regular hexagonal packing geometry, while the other phase corresponds 
 to a network with many soft deformation modes, where the hexagonal 
 packing looses stability.
Here, we develop a general framework to study the stability of tissues 
subject to cell activity and externally applied stresses. Neither cell division nor cell rearrangements are considered. This is the case of some experiments \cite{zallen2004cell, harris2012} and previous analytical calculations \cite{staple2010mechanics, merzouki2016, nestor2018}. Also, topological events are non-linear and, therefore, they are not relevant to describe the emergence of the instabilities.
We show that for a large region of the parameter space, if in large portions of the tissue the cells modify their activity or it is subject to external stresses, the whole tissue becomes unstable in the form of long-wavelength deformations that 
couple pure-shear or deviatoric modes, with rotation and expansion 
modes. These instabilities  differ from those that take place in passive foams \cite{cohen2013flow,spencer2017vertex}, because they are triggered by the cellular activity.

The organization of the paper is as follows. In Sec.~\ref{sec.activity} we present the general analysis of the instabilities that appear in a confluent tissue, focusing in the case of cellular activity. The analytical method for regular tissues and the simulations for irregular ones are described and compared. Section \ref{sec.prestress} considers the case of tissues subject to external pre-stresses. In Sec.~\ref{sec.analysisnondiag} we discuss the case of general anisotropic pre-stresses, which need a more detailed analysis. Our conclusions and a discussion are presented in Sec.~\ref{sec.discussion}. Finally, the Appendices give technical details.

%%%%%%
\section{Tissue under cell activity} \label{sec.activity}
For the vertex model, the elastic 
coefficients $K_A$ and $K_P$ are assumed to be positive, and penalize deviations 
from the reference areas and perimeters, while there is no restriction on the sign of 
$J$, as has been discussed in the literature~
\cite{fletcher2014vertex,jessica2017quantitative}. 
As a first case, where analytical results can be obtained, we consider a regular tissue 
composed of $N$ identical hexagonal cells of side $a$, for which $A_{0c}=3\sqrt{3}a^2/2$ and 
$P_{0c}=6a$, for all cells $c$. 
Cell activity can generate stresses that tend to deform the tissue. For example,
sudden changes in the actomyosin activity in the cell border  can be modeled  as a 
modification of the equilibrium perimeters, $P_{0c} \rightarrow \left(1+\lambda_P \right) P_{0c}$ (with 
$\lambda_P>0$ for expansions and $\lambda_P<0$ for contractions). Similarly, a change in the 
actomyosin activity in the medioapical side of the cells imply changes in the equilibrium cell 
areas, $A_{0c} \rightarrow \left(1+\lambda_A \right) A_{0c}$. 

As a first case, we consider homogeneous modifications of the tissue (uniform $\lambda_P$ and $\lambda_A$), modeling large portions of the tissue that change as in Ref.~\cite{harris2012},  and we 
investigate the stability and rigidity of this tissue, allowing it to fluctuate. The vertex 
positions are now given by $\left(I+\epsilon U\right) \VEC r^{[0]}_i$, where $\epsilon\ll 
1$, and $U$ a general  $2\times2$ matrix of components $u_{ik}$, characterizing the 
fluctuations. Computing contributions up to $O(\epsilon^2)$, the energy of the tissue 
may be written as
$E = \sum_{i=0}^2\epsilon^{i}\left(E_A^{(i)} + E_P^{(i)} + E_J^{(i)}\right)$,
where the superscripts represent the order of each term in the expansion, and 
$E_A$, $E_P$ and $E_J$ are the contributions proportional to $K_A$, $K_P$ and $J
$, respectively. The full expressions are given in Appendix \ref{ap.energy.activity}.

The stress tensor is $\sigma_{ik} = \derpar{E}{u_{ik}}$. 
It has a zeroth order contribution derived from $E^{(1)}$, $\sigma_{ik}^{(0)}= 2 \sqrt{2} \hat E  \left(\frac{2}{9}j  -\lambda_A - \frac{8}{3}p\lambda_P\right)\delta_{ik}$ that represents the total stress, with passive and active contributions, needed to maintain the  deformed configuration. Here, we defined the energy  scale $\hat{E} =  {N K_A} A_{0}^2/2$ and the dimensionless parameters $p=K_P/(a^2K_A)$ and $j=J/(a^3 K_A)$, which are the ratios between the characteristic time of the surface elasticity and the ones related to the perimeter and adhesion elasticity, respectively.

For general fluctuations, $U$ can be expanded in Fourier modes. When computing the 
total energy of the tissue, the linear terms in $\varepsilon$ cancel by spatial 
integration, leaving only the reference energy and the quadratic terms in the 
fluctuations. In physical terms, the linear contribution is eliminated by the application of 
a uniform external stress $\sigma_{ik}^{(0)}$ by other tissues that act as a frame, 
imposing rigid boundary conditions.  Furthermore, in the limit of small wavevectors $
\VEC k$, the dominant contribution comes from the case of homogeneous $U$, plus 
small corrections proportional to $k^2$, which we  neglect henceforth. Hence, to 
analyze the stability of the tissue under long wavelength fluctuations, we have to  
determine whether the quadratic form for homogeneous $U$ is positive definite. Expressing 
$U$ as a linear combination of four basic deformation modes,
\begin{align}
U_1 &=\begin{pmatrix}
-1 & 0\\
0 & 1
\end{pmatrix} \text{[deviatoric]},  &
U_2 &= \begin{pmatrix}
0 & 1\\
1 & 0
\end{pmatrix} \text{[pure shear]}, \label{eq.modes}\\
U_3 &= \begin{pmatrix}
0 & -1\\
1 & 0
\end{pmatrix} \text{[rotation]},&
U_4 &= \begin{pmatrix}
1 & 0\\
0 & 1
\end{pmatrix} \text{[expansion]},\nonumber
\end{align}
as $U=\sum_{i=1}^4  v_i U_i$, the energy can be expanded as $E^{(2)}=\hat{E} \sum_{i,j=1}^4 \mu_{ij} v_iv_j$.
In the case where the deformation is due to cell  activity,  the $\mu$-matrix is diagonal with 
\begin{align}
\mu_{11}&=\mu_{22}=
\frac{j}{9} + \lambda_A -\frac{4 p \lambda_P}{3},  \\
\mu_{33}&=
\frac{2j}{9} - \lambda_A-\frac{8 p\lambda_P}{3}, 
\mu_{44}=2 + \frac{8p}{3} - \lambda_A,
\end{align}
where we used the expressions of Appendix \ref{ap.energy.activity}.
The deformation modes $U_1$ and $U_2$ are both shears, although in  different 
directions. Consequently, their eigenvalues, which are associated to the shear 
modulus, are equal. Negative values of the diagonal terms signal the development 
of an instability of the 
corresponding mode, in a single cell description. For example, large positive values of 
$\lambda_A$ (cell expansion), would give rise to unstable rotation and expansion modes, while 
for large negative values of $\lambda_A$ (cell compression), the deviatoric and pure shear 
modes become unstable. 

\begin{figure}[htb]
 \includegraphics[width=.95\linewidth]{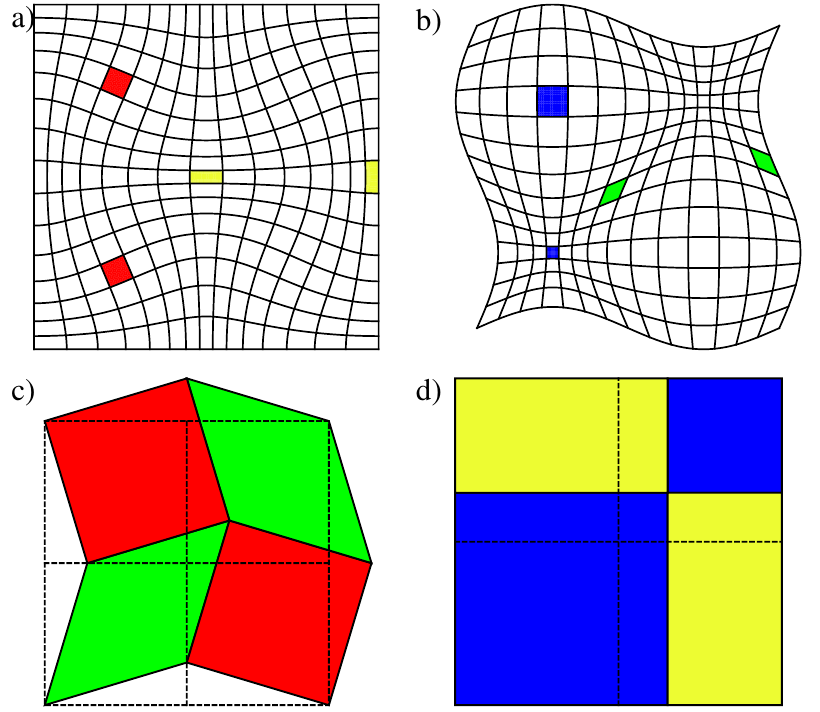} 
\caption{Representation of tissue fluctuations where cells subject to different deformation modes coexist. Long wavelength fluctuations a) $x'=x+ \epsilon\sin(2\pi x/L)\cos(2\pi y/L)$ and $y'=y-\epsilon\sin(2\pi y/L)\cos(2\pi x/L)$, and b) $x'=x+ \epsilon\cos(2\pi x/L)\sin(2\pi y/L)$ and $y'=y+\epsilon\cos(2\pi y/L)\sin(2\pi x/L)$. 
c) and d) fluctuations where the boundaries between neighbor cells move inside a supercell. Cells with well defined deformation modes are colored: yellow for deviatoric, green for pure shear, red for rotation, and blue for expansion. For simplicity, square cells are used in the presentation.}
\label{fig.modecoupling}
\end{figure}

At a tissue level, however, due to the confluent property, pure 
modes are not allowed. 
Indeed, consider for example the Fourier mode where the new vertex positions 
are given by 
$x'=x+ \epsilon\sin(2\pi x/L)\cos(2\pi y/L)$ and $y'=y-\epsilon\sin(2\pi y/L)\cos(2\pi x/L)$, 
shown in Fig.~\ref{fig.modecoupling}a. 
Depending on the position, some cells experience deviatoric deformations (in yellow), while others rotate (in red). 
Similarly, for the Fourier mode 
$x'=x+ \epsilon\cos(2\pi x/L)\sin(2\pi y/L)$ and $y'=y+\epsilon\cos(2\pi y/L)\sin(2\pi x/L)$, 
shown in Fig.~\ref{fig.modecoupling}b, 
pure shear modes (in green) coexist with expansion modes (in blue). Simple uniaxial deformations with a sinusoidal amplitudes also couple the deviatoric and expansion modes.
Complementary to the long wavelength fluctuations, it is possible that the boundaries between neighboring cells move inside a supercell (analogous to optical phonons in solids) as shown in Figs.~\ref{fig.modecoupling}c and d. Again, different modes coexist. The confluent property with the periodic boundary conditions frustrate the emergence of pure deformation modes. The use of fixed boundary conditions leads to the same frustration.

This unavoidable coexistence of modes implies that even though a deformation mode may seem to be unstable at the cell level, the total energy of the tissue should be computed as the sum of the different contributions that, at the end, may result to be positive definite. A detailed study of the stability of a tissue that considers the coexistence of modes is  given in Section \ref{sec.analysisnondiag}. We provide here a qualitative argument to obtain the stability limit from the behavior of individual cells. As the deviatoric and pure shear modes share the same value in the $\mu$-matrix, the total energy of the tissue fluctuations shown in Figs.~\ref{fig.modecoupling}a and c are equal, with a prefactor equal to $\mu_{11}+\mu_{33}=j/3-4p\lambda_P$. An instability is hence predicted to develop for $\lambda_P>j/(12p)$. Notably, when $\lambda_A=0$, the instability is predicted to take place when the shear modulus (i.e.\ $\mu_{11}$ or $\mu_{22}$) vanishes, as was observed in Ref.~\cite{staple2010mechanics}. However, when the target area has changed ($\lambda_A\neq0$), the vanishing of the shear modulus does not signal the development of unstable modes.

To validate the predictions in actual situations, we simulate both regular and irregular tissues. 
Regular hexagonal tissues are made of $N=3000$ cells arranged in a box of size $L_x = 50 \sqrt{3}a$ and $L_y=90a$ with periodic boundary conditions. In order to avoid artificial effects due to the lattice perfections, a Gaussian noise is added to all the vertex positions in both directions, with standard deviation $0.1a$.
 Irregular tissues are built as Voronoi cells, where the positions of $N=3000$ center points are generated by a Montecarlo simulation of hard disks in a box of equal size as for the regular tissue. The  diameter of the disks govern the degree of dispersion of the cells. We consider an area fraction $\phi=0.71$, below the freezing transition, to obtain a reproducible disordered tessellation with moderate dispersion in cell sizes. The irregular tissues are made of polygons of different sizes and number of sides, implying variance in the equilibrium areas and perimeters, $A_{0c}$ and $P_{0c}$. The deviatoric and pure shear modes manifest in the elongation of cells, which we characterize by the flattening parameter $c=(a-b)/(a+b)$, computed for each cell in terms of its principal semiaxis $a$ and $b$,
 calculated as the square root of the eigenvalues of the texture matrix
$M_c =\frac{1}{n_c}\sum_{i\in c}\left( \VEC{r}_i - \VEC{r}_c\right)\otimes \left( \VEC{r}_i - \VEC{r}_c\right),$
where the sum is over the $n_c$ vertices conforming the cell, with positions $\VEC{r}_i$, and $\VEC{r}_c$ is the center of the cell.
Simulations are performed solving numerically the equations of motion \eqref{eq.variationaldyn}, which are worked out in the Appendix \ref{app.eqmotion} [Eqs.~\eqref{eq.eq_mot}, \eqref{eq.vA}, \eqref{eq.vP}, and \eqref{eq.vJ}]. The differential equations are integrated using the Euler integration method, for various values of $K_P$ and $J$, fixing units such that $K_A=1$ and $a=1$. The time step was fixed to $dt=0.005$ and we study the system up to $t=0.5$.

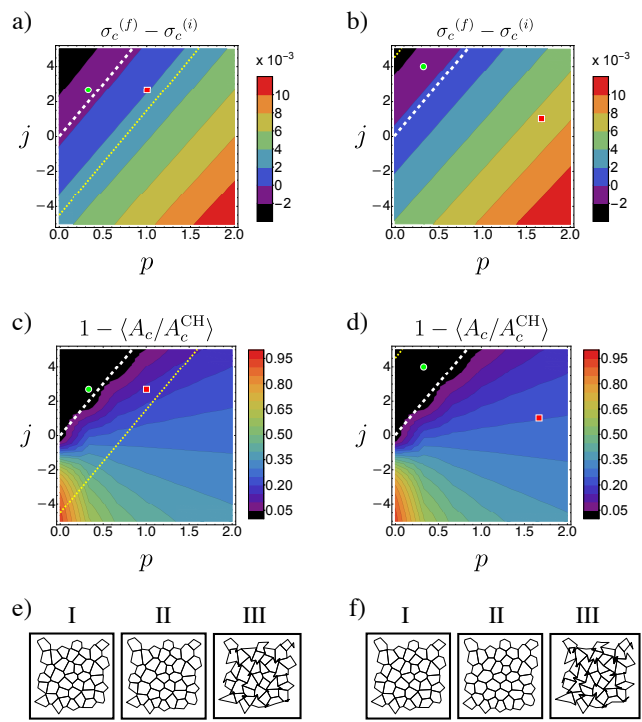
\begin{figure}[ht] 
 \includegraphics[width=\linewidth]{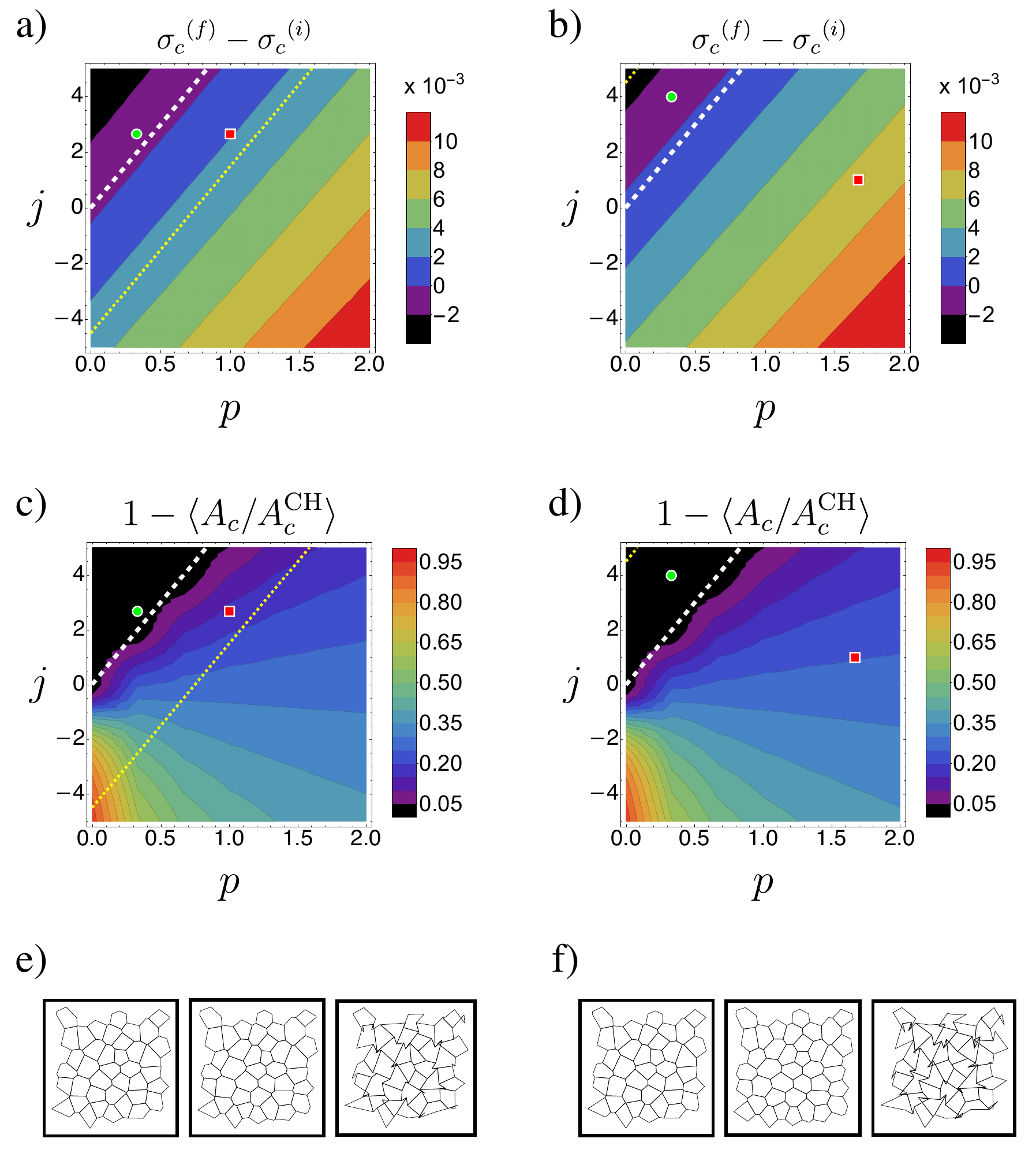} 
 \begin{picture}(0,0)(0,0)
\put(-98,55){I}
\put(-65,55){II}
\put(-32,55){III}
\put(28,55){I}
\put(61,55){II}
\put(94,55){III}
 \end{picture}\vspace{-0.2cm}
 \caption{Tissue instabilities obtained in simulations of $N=3000$ irregular cells under the action of cell activity: modification of the equilibrium perimeter with $\lambda_P=1/2$ and the equilibrium area with $\lambda_A=1/2$ (left) and $\lambda_A=-1/2$ (right).  Top: change of the standard deviation of  the flattening parameter after a short time, $t=0.025$. Negative values indicate cells become more uniform. Middle: one minus the mean value of the area of each cell divided by the area of the associated convex hull, after a longer time, $t=0.5$. Units are fixed such that $K_A=1$ and $a=1$. See Appendix \ref{app.timescales} for an analysis of the relevant time scales, justifying the election of the observation times.  The thick white line and the thin yellow line are the analytical curves obtained when assuming or neglecting coupling of modes, respectively. Instabilities are predicted to the right of the lines. Note that in panels b) and d), the thin yellow line is close to the top-left corner.
 Bottom:  Examples of a section of an irregular tissue for each case of cell activity,  indicating  (I) the initial configuration at t = 0, and the final configurations at t = 0.5, for the cases of the (II) green-disk/stable and (III) red-square/unstable markers. The results are the average of six different irregular tissues, generated with the same parameters.}
 \label{fig.resultados_coupledmodes}
 \end{figure}

The change of the standard deviation of the flattening parameter after few time steps for fixed positive perimeter change $\lambda_P=1/2$, considering $\lambda_A=\pm 1/2$, displays an important increase precisely where the instability is predicted (Figs.~\ref{fig.resultados_coupledmodes}a and b). The chosen values of $\lambda_{A,P}$ are consistent in the order of magnitude with experiments using laser ablation and biochemical perturbations \cite{farhadifar2007influence,rauzi2008nature,harris2012}. For larger times, an important fraction of the polygons become non-convex as a consequence of the instability (Figs.~\ref{fig.resultados_coupledmodes}e and f). The non-linear dynamics does not saturate the instability and, from a practical point of view, this implies that the vertex model ceases to be a valid description of tissues when these instabilities develop. Nevertheless, the non-convexity can be used as a proxy of the instability and, for a continuous quantification, one minus the mean value of the area of each cell divided by the area of the respective convex hull is presented in Figs.~\ref{fig.resultados_coupledmodes}c and d. For convex polygons, this order parameter vanishes, while positive values indicate that non-convex polygons appear. The agreement with the analytical prediction is excellent, both when regular and irregular tissues are simulated. A comparison between regular and irregular tissues is presented in the Appendix  \ref{app.comparison}, showing that the instability takes place for the same parameters and the values of the observables agree. 
Importantly, the line at which the shear modulus vanishes ---obtained when neglecting the coupling of modes--- fails to predict the instability for all tissues (Figs.~\ref{fig.resultados_coupledmodes}, \ref{fig.simP0_1}, and \ref{fig.simP0_2}).

For the cases shown in Figs.~\ref{fig.modecoupling}b and d, the energy for the tissue has a prefactor that becomes negative when $p\lambda_P>3/2+2p+j/12$, requiring  an extremely large increase of the equilibrium perimeter, except if $j$ is negative. Consequently, these modes are hardly seen and are hidden by other more unstable modes. 

For cells of equal equilibrium area and complete contraction of the perimeter ($\lambda_P=-1)$, the transition line in Refs.~\cite{farhadifar2007influence, staple2010mechanics} is reproduced. An important difference with their work is 
the use of a fixed size box in simulations, generating at long times non-convex polygons instead of soft networks.

\section{Tissue under pre-stress} \label{sec.prestress}
In addition to cellular activity, the tissue can be subject to a pre-stress generated by  the action of neighboring cells or tissues, fixed boundary conditions, an actomyosin network, or the drag by another expanding tissue located in an adjacent layer, causing it to get pre-deformed.
 To model a pre-stressed tissue, we perform an affine transformation by changing the vertices positions as $\VEC r^{[0]}_i \rightarrow
\Lambda\VEC r^{[0]}_i$, where $\Lambda$ is the $2\times2$ matrix associated to the pre-deformation.
Adding fluctuations, the vertex positions are now given by $\left(I+\epsilon U\right)  \Lambda\VEC r^{[0]}_i$.

As for the cell activity, we consider homogeneous deformations of the tissue (uniform $\Lambda$) and perturbations $U$ in the small wavevector limit, and we analyze first the different deformation modes independently, without dealing with their coupling. 
For an hexagonal cell, it is found that $E_A^{(2)} = \hat{E}\left[ \mathrm{det}(\Lambda)^2 \mathrm{tr}(U)^2+ 2\mathrm{det}(\Lambda)\left(\mathrm{det}(\Lambda)-1\right)\mathrm{det}(U)\right]$.
The expressions for $E_P^{(2)}$ and $E_J^{(2)}$ are  more involved but numerically it is found that they are always positive definite for all pre-deformations, when $K_P$ and $J$ are positive (see Appendix \ref{ap.energy.stress} for the full expressions). We conclude, then, that negative $J$ could give rise to instabilities for any pre-strain. 
The case of $E_A^{(2)}$ requires more analysis.
From  the expression for $E_A^{(2)}$, it is found that fluctuations with $\mathrm{det}(U )=0$  are always stable. Using  the expansion $U=\sum_{i=1}^4  v_i U_i$, $E_A^{(2)}$ is diagonal with elements  $\mu_{A11} = \mu_{A22} = -\mu_{A33} =-\bar{\lambda}$, with $\bar\lambda = \frac{27}{8}[ \mathrm{det}(\Lambda) - 1]\mathrm{det}(\Lambda)$, and $\mu_{A44} = \frac{81}{8}\mathrm{det}(\Lambda)\left[\mathrm{det}(\Lambda) -1/3 \right]$. Note that whenever $\mathrm{det}(\Lambda)\neq0$, either $\mu_{A11,A22}$ or $\mu_{A33}$ are negative, giving rise to possible unstable modes. When $\mathrm{det}(\Lambda)>1$ (for example, under a pre-expansion), $\mu_{A11,A22}$ are negative and the deviatoric and pure shear modes may be unstable. Also, when $0<\mathrm{det}(\Lambda) < 1$ (for example, under a compression pre-deformation), $\mu_{A33}$ is negative and the rotation mode may be unstable. To fully determine the stability, we must consider the perimeter and edge contributions to the energy, as well as the mode couplings.

\begin{figure}[ht] 
 \includegraphics[width=\linewidth]{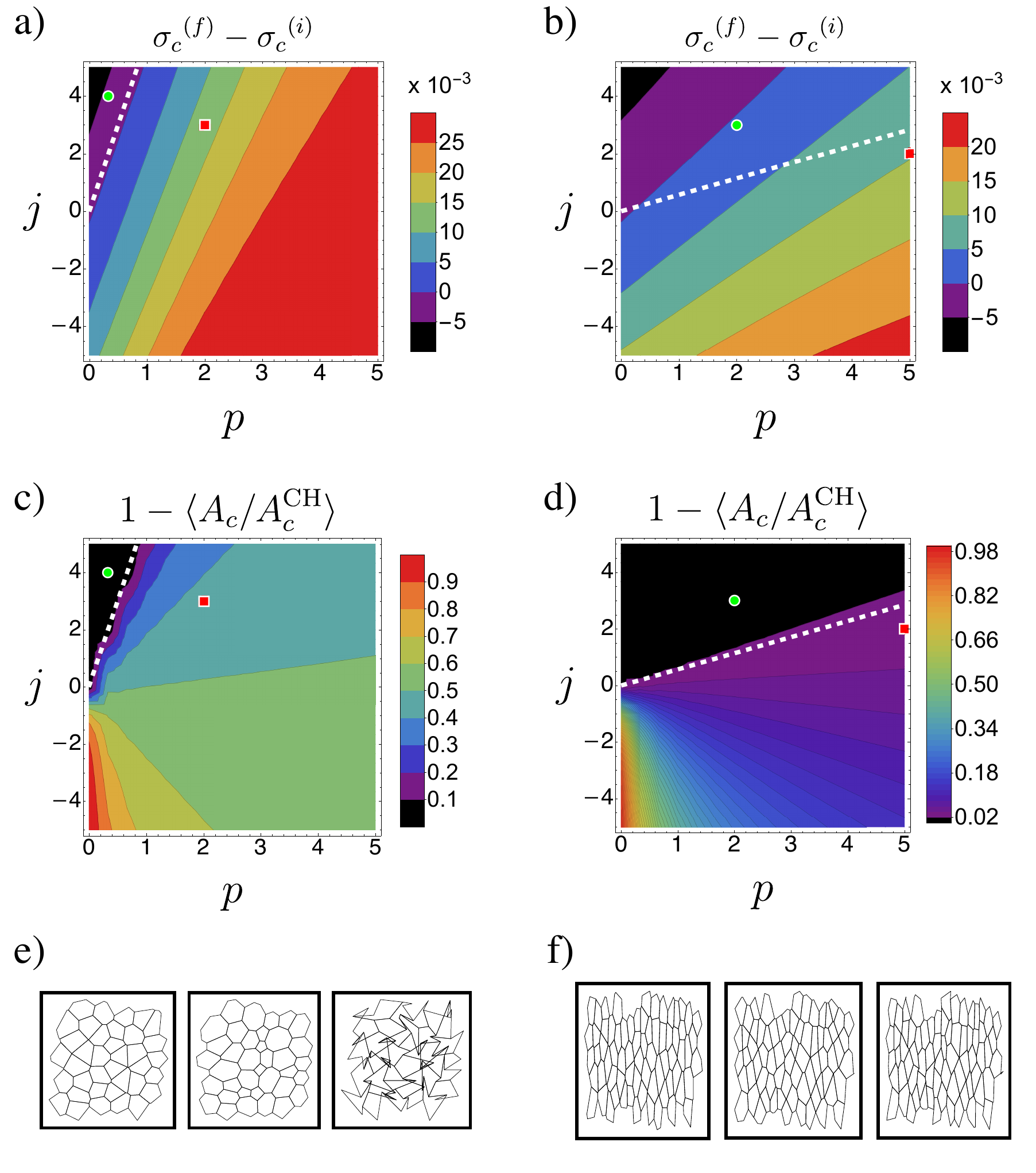} 
 \begin{picture}(0,0)(0,0)
\put(-98,57){I}
\put(-65,57){II}
\put(-32,57){III}
\put(28,57){I}
\put(61,57){II}
\put(94,57){III}
 \end{picture}\vspace{-0.2cm}
\caption{Tissue instabilities obtained in simulations of $N=3000$ irregular cells 
in tissues under $50\%$ isotropic contraction (left), and under $60\%$ 
horizontal contraction plus $40\%$ vertical expansion (right). Same 
representation as in Fig.~\ref{fig.resultados_coupledmodes}.}
\label{fig.resultados2}
\end{figure}
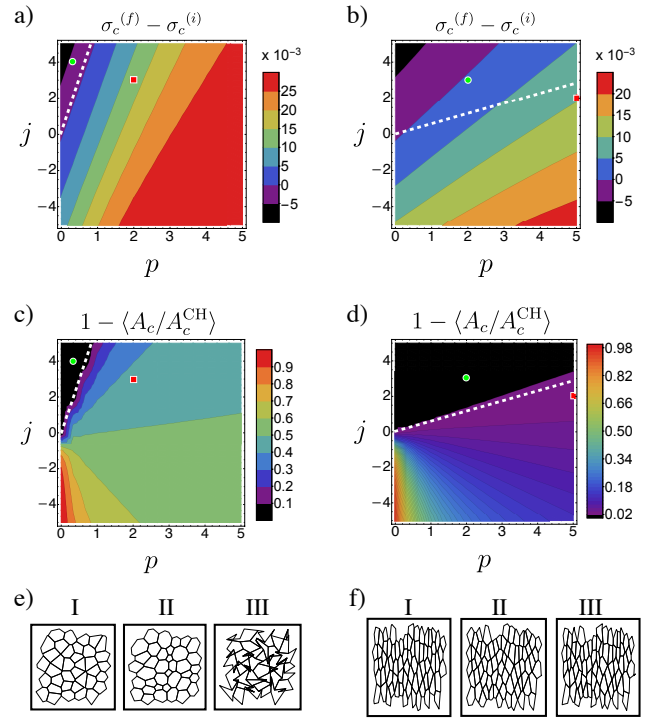

For isotropic pre-strain $\Lambda=(1+h)I$ ($h>0$ for expansions and $-1<h<0$ for compressions), the complete $\mu$-matrix is diagonal, with
\begin{align}
\mu_{11}&=\mu_{22}=(1+h) (-2 h - 3 h^2 - h^3 + 4 h p/3 + j/9), \\
\mu_{33}&=(1+h) (2 h + 3 h^2 + h^3 + 8 h p/3 + 2 j/9), \\
\mu_{44}&=(1+h)(2 + 8 h + 9 h^2 + 3 h^3 + 8p/3 + 8 hp/3).
\end{align}
The stability of the relevant global mode is, therefore, described by $\mu_{11}+\mu_{33}=(1+h)(4 h p + j/3)$, which can become negative for a wide range of parameters when the tissue is under compression.
Simulations are performed,  using the  methods described  in Section \ref{sec.activity}, for an isotropic compression of 50\%. Figure \ref{fig.resultados2}-left shows an excellent agreement with the analytical calculations that predict the instability line at $j=6p$. Again, the instability manifests in an increase of the eccentricity and, at longer times, the appearance of non-convex polygons.

\section{Anisotropic pre-stresses}\label{sec.analysisnondiag}

Finally, {\it in vivo} or {\it in vitro} tissues are in general subject to anisotropic 
external deformations \cite{mao2011planar,rauzi2008nature,leptin1990cell}, causing the $\mu$-matrix to be non-diagonal. The relevant 
global modes are obtained as follows. For an extended tissue, the 
fluctuation is expanded in Fourier modes: ${\VEC r'} ={\VEC r}+ \sum_{\VEC k}{\VEC a}_{\VEC k}e^{i\VEC k
\cdot\VEC r}$.
From the Jacobian of this transformation, the local deformation matrix is computed as  $u_{\alpha \beta}(x,y)=ik_{\alpha} a_{\VEC k \beta}e^{i\VEC k
\cdot\VEC r}$. Expanding it as $U(x,y)=\sum_{i=1}^4  v_i(x,y) U_i$, a local energy density is obtained, $e(x,y)=(\hat{E}/L^2) \sum_{i,j=1}^4 \mu_{ij} v_i(x,y)v_j(x,y)$. 
Finally, the total energy of the tissue is
\begin{align}
E=\int dx\, dy\, e(x,y) =\sum_{\VEC k} \sum_{\alpha,\beta=1}^2 k^2 e_{\alpha\beta}(\HAT k)  a_{\VEC k \alpha} a^{*}_{\VEC k\beta},
\end{align}
where we used that  $v_i(x,y)$ are linear combinations of the Fourier coefficients ${\VEC a}_{\VEC k}$ and that the Fourier modes decouple if the tissue is homogeneous on the large scale. 
The matrix $e_{\alpha\beta}$ is a $2\times2$ matrix with real coefficients.
\begin{align}
 e_{11} = & \frac{1}{4}\big\{(\mu_{13} +\mu_{24})\sin2\theta  +  (\mu_{11} - 2 \mu_{14} + \mu_{44})\cos^2 \theta\nonumber\\ 
&+    [ (\mu_{22} - 2 \mu_{23} + \mu_{33})\sin\theta  - 2(\mu_{12} + \mu_{34})  \cos\theta]\sin\theta  \big\}, \label{eq.e11}\\
   e_{12} = &  e_{21} = \frac{1}{8} \big\{2 [-\mu_{13} + \mu_{24} +  (-\mu_{12} + \mu_{34})\cos2 \theta] \nonumber\\
   &+   (-\mu_{11} + \mu_{22} - \mu_{33} + \mu_{44})\sin2 \theta\big\},\label{eq.e12}\\
    e_{22} = & \frac{1}{4} \big\{ (\mu_{22} + 2 \mu_{23} + \mu_{33})\cos^2\theta +  (\mu_{11}  + 2 \mu_{14} + \mu_{44})\sin^2\theta  \nonumber \\
   &     +   (\mu_{12} + \mu_{13} + \mu_{24} + \mu_{34}) \sin2 \theta
   \big\}\label{eq.e22}.
\end{align}
where we used that the $\mu$-matrix is symmetric.
The stability of the tissue, considering the confluent and periodic conditions, is then obtained from the eigenvalues of the $e$-matrix, which depend only on the direction $\HAT{k}$ of the wavevector. If at least one eigenvalue is negative, the tissue develop long wavelength instabilities.
When the $\mu$-matrix is diagonal, and using that $\mu_{11}=\mu_{22}$, it is found that the eigenvalues of $e_{\alpha\beta}$ do not depend on $\theta$ and  they are given by $\frac{1}{4} \left(\mu_{11} + \mu_{33}\right)$ and $\frac{1}{4} \left(\mu_{11} + \mu_{44}\right)$, which corroborates the simple analysis for the coupling of modes  described in Section \ref{sec.activity}.

\begin{figure}[htb] 
    \centering
    \includegraphics[width=0.98\linewidth]{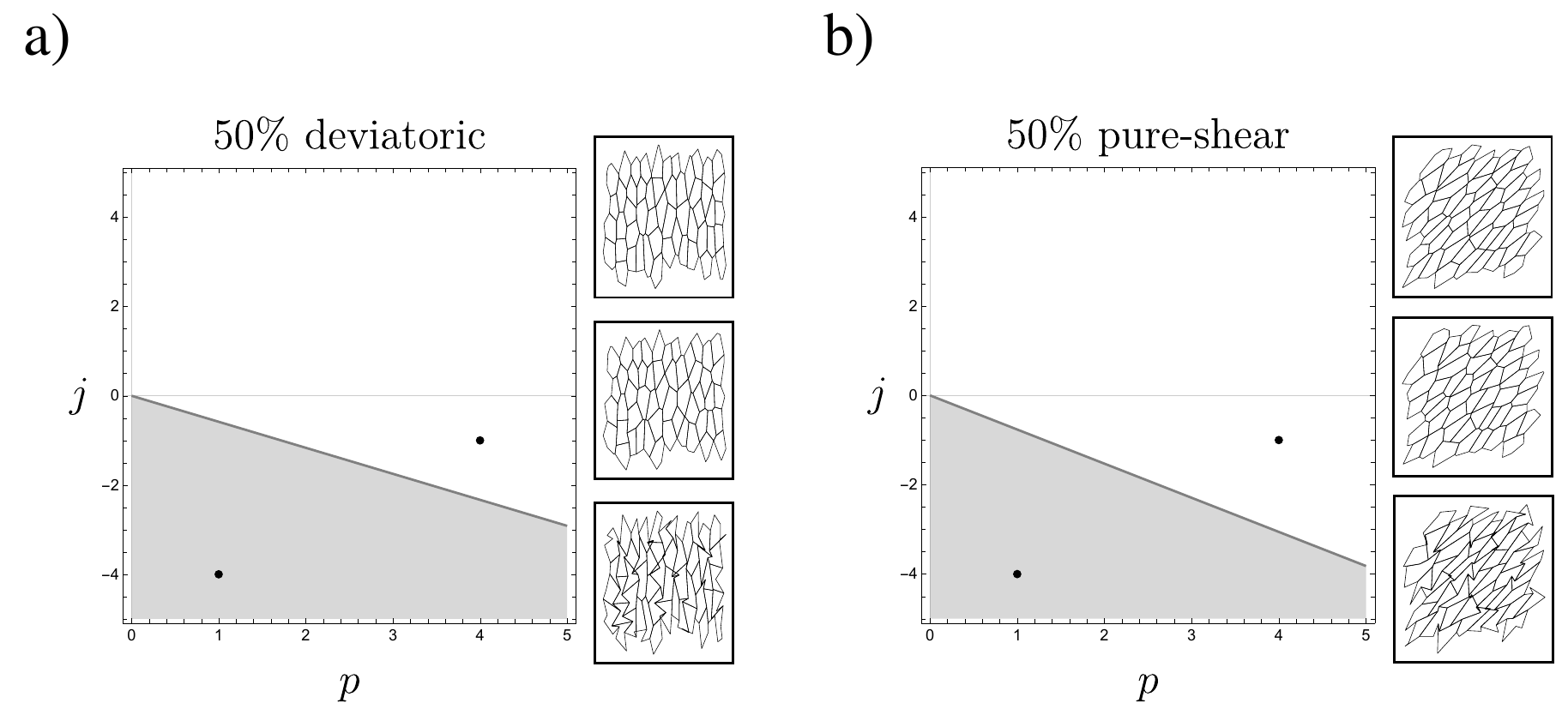}
 \caption{  Transition line at which the minimum eigenvalue of the associated $e$-matrix changes its sign, for a tissue under  a)  $50\%$ deviatoric pre-stress and b) $50\%$ pure shear pre-stress. The gray areas correspond to the unstable part of the parameter space. Sections of an irregular tissue are shown for each case, indicating the initial configuration (top), at $t=0$, and the final configurations, at $t=0.5$,  for the cases of the marked black dots, one stable (middle) and other unstable (bottom).}
 \label{fig.act_otros}
 \end{figure}

Anisotropic pre-deformations generate non-diagonal $\mu$-matrices, for which some examples are given in the Appendix~\ref{app.nondiagonal}. Figure~\ref{fig.resultados2}-right presents the comparison between simulations and the prediction of the instability using the eigenvalues of the associated $e$-matrix for a tissue under $60\%$ horizontal contraction plus $40\%$ vertical expansion. The agreement is again excellent when the non-convexity proxy is used. The flattening parameter does not signal the instability because, for this case there is no manifestation in the change of ellipticity as a result of the coupling of all modes. Finally, Fig.~\ref{fig.act_otros} shows the results for a tissue that is subject to a pure deviatoric stress or to a pure shear stress. 

\section{Discussion} \label{sec.discussion}
Our analysis shows that stressed tissues described by the two-dimensional  vertex model present instabilities in which the cells deform to increase their ellipticity, to later become non-convex. These stresses can be generated by the cellular activity when the actin ring on the perimeter of the cells changes its size or they can be external, when the tissue is pre-stressed. In any of these cases the tissue is unstable for a wide range of the  model parameters.

The presence of the predicted instabilities is a stringent test of the vertex model to describe biological tissues, which under many conditions are subject to internal and external stresses. For example, in developing tissues, 
processes like invaginations, cell extrusion and division generate stresses. Uniaxial pulling can be generated by other tissues~\cite{etournay2015interplay} or driven experimentally~\cite{koshihara2010effect,nestor2018,harris2012}. Also, biochemical signals can alter in large regions the activity of the tissue~\cite{harris2012}. These and other configurations, with different external stresses, should be investigated to verify if the predicted instabilities take place and if they can act as seeds to instabilities in developing tissues.
In the mechanobiological approach, forces and instabilities launch the tissue transformations during development that are necessary to generate structures and organs~\cite{li2012mechanics, nelson2016buckling}. If the vertex or similar models correctly describe the tissue dynamics, internal or external stresses can trigger the instabilities described in this letter, which can initiate  tissue transformation processes.

In this letter we restricted the analysis to two-dimensional planar dynamics. Further studies are needed to analyze how  the deformation modes couple with motion in the third dimension when the planar restriction is removed. For  example, buckling instabilities generating wrinkles, could relax stresses instead of generating non-convex polygons.

\acknowledgments
This research was supported by the Franco-Chilean EcosSud Collaborative Program C16E03, 
the Fondecyt Grant No.\ 1180791 and  the Millennium Nucleus Physics of Active Matter of ANID (Chile).

\appendix

  \section{Energy expressions for fluctuating tissues} \label{ap.energy}
 
For the analytic calculations, we consider a regular tissue composed of $N$ identical regular hexagonal cells of side $a$, for which the preferred cell area and perimeter for all cells are  $A_{0c}=3\sqrt{3}a^2/2$ and $P_{0c}=6a$, respectively. 

  \subsection{Tissue under cell activity}\label{ap.energy.activity}

Cell activity is included as homogeneous modifications of the equilibrium perimeters, 
$P_{0c} \rightarrow \left(1+\lambda_P\right) P_{0c}$ and equilibrium areas $A_{0c} \rightarrow \left(1+\lambda_A\right) A_{0c}$,  
with $\lambda_P,\lambda_A>0$ for expansions and $\lambda_P,\lambda_A<0$ for contractions.

We define $A_c^{(1)}$ as the area of the cell $c$ with fluctuations characterized by the matrix $U$,
\begin{align}
A_c^{(1)} = \left(1+\epsilon \mathrm{tr}\left(U\right) + \epsilon^2 \mathrm{det}\left(U\right)\right)A_{0c}.
\label{eq.A1}
\end{align}
Then, when considering an activity modulated by $\lambda_A$, the term of the energy proportional to $K_A$ is given by
\begin{align}
E_A = &\sum_c \frac{K_A}{2}\left[ A_c^{(1)} - \left(1+\lambda_A\right)A_{0c} \right]^2 \nonumber, \\
=&\sum_c \frac{K_A}{2}A_{0c}^2\left[ -\lambda_A + \epsilon \mathrm{tr}\left(U\right) + \epsilon^2 \mathrm{det}\left(U\right)\right]^2.
\label{eq.EA2}
\end{align}
Hence, the zeroth, first, and second order terms of $E_A$ are 
\begin{align} 
E_{A}^{(0)} & = \sum_c \frac{K_A}{2} A_{0c}^2  \lambda_A^2, \label{AreaEnergy02}\\
E_{A}^{(1)} & = -\sum_c K_A A_{0c}^2 \mathrm{tr}\left(U\right) \lambda_A, \label{AreaEnergy11}\\
E_{A}^{(2)} & = \sum_c \frac{K_A}{2} A_{0c}^2 \left[\mathrm{tr}\left(U\right)^2-  2\mathrm{det}\left(U\right)\lambda_A\right]. \label{AreaEnergy21} 
\end{align}

We define $P_c^{(1)}$ as the perimeter of the cell $c$ with fluctuations characterized by the matrix $U$,
\begin{multline}
P_c^{(1)} = \left[1+\frac{1}{2}\epsilon \mathrm{tr}\left(U\right) + \frac{1}{8}\epsilon^2 \mathrm{det}\left(U\right)\right.\\
\left.+  \frac{3}{16}\epsilon^2 \mathrm{tr}\left(U^{T}U\right)-\frac{1}{8}\epsilon^2 \mathrm{tr}\left(U\right)^2\right]P_{0c}.
\label{eq.P1}
\end{multline}
Then, when considering an activity modulated by $\lambda_P$, the term of the energy proportional to $K_P$ is given by
\begin{align}
E_P = &\sum_c \frac{K_P}{2}\left[ P_c^{(1)} - \left(1+\lambda_P\right)P_{0c} \right]^2\nonumber, \\
=&\sum_c \frac{K_P}{2}P_{0c}^2\left[ -\lambda_P+ \frac{1}{2}\epsilon \mathrm{tr}\left(U\right) + \frac{1}{8}\epsilon^2 \mathrm{det}\left(U\right)\right.\\
& \left.+  \frac{3}{16}\epsilon^2 \mathrm{tr}\left(U^{T}U\right) -\frac{1}{8}\epsilon^2 \mathrm{tr}\left(U\right)^2\right]^2.
\label{eq.EP1}
\end{align}
The zeroth, first, and second order terms of $E_P$ are therefore given by
\begin{align} 
E_{P}^{(0)} & = \sum_c \frac{K_P}{2} P_{0c}^2  \lambda_P^2, \label{PerEnergy0} \\
E_{P}^{(1)} & = -\sum_c \frac{K_P}{2} P_{0c}^2 \mathrm{tr}\left(U\right) \lambda_P, \label{PerEnergy1} \\
E_{P}^{(2)} & = \sum_c   \frac{K_P}{8}P_{0c}^2 \left[  \left(1+\lambda_P\right) \mathrm{tr}\left(U\right)^2 -\lambda_P   \mathrm{det}\left(U\right)-\frac{3}{2}\lambda_P  \mathrm{tr}\left(U^{T}U\right)  \right]. \label{PerEnergy2}
\end{align}

Finally, the adhesion contribution to the energy is
\begin{align}
E_J = &\sum_c \frac{J}{2} P_c^{(1)}, \label{eq.EJ1}
\end{align}
where $P_c{(1)}$ is given in Eq.~\eqref{eq.P1},
As a result, the zeroth, first, and second order terms of $E_J$ are given by
\begin{align} 
E_{J}^{(0)} & = \sum_c   \frac{J}{4} P_{0c}, \label{JEnergy01} \\
E_{J}^{(1)} & = \sum_c   \frac{J}{4} P_{0c} \mathrm{tr}\left(U\right), \label{JEnergy11} \\
E_{J}^{(2)} & = \sum_c \frac{J}{16} P_{0c} \left[  \mathrm{det}\left(U\right) +  \frac{3}{2} \mathrm{tr}\left(U^{T}U\right) -\mathrm{tr}\left(U\right)^2\right].  \label{JEnergy21}
\end{align}

Eqs.~\eqref{eq.A1} and \eqref{eq.P1} can be obtained using Mathematica.

 \subsection{Tissue under stress} \label{ap.energy.stress}

Now, we study the same energy contributions, but when the tissue is subject to a homogeneous strain, such that all the vertices change their position as $\VEC r^{[0]}_i \rightarrow
\Lambda\VEC r^{[0]}_i$, where $\Lambda$ is a $2\times2$ matrix that gives account of the pre-deformation. 

In a similar way as in the previous section we can define  $A_c^{(1)}$ and $P_c^{(1)}$, representing the area and perimeter of the cell $c$, that was initially a regular hexagon with area $A_{0c}$ and perimeter $P_{0c}$, which is now subject to a given strain characterized by the matrix $\Lambda$. Then, we define $A_c^{(2)}$ and $P_c^{(2)}$ as the values when we allow fluctuations, modulated by the matrix $U$, in the system.
\begin{align}
A_c^{(1)} = & \mathrm{det}\left(\Lambda \right)A_{0c}, \label{eq.A1_2}\\
A_c^{(2)} = & \left[1+\epsilon \mathrm{tr}\left(U\right) + \epsilon^2 \mathrm{det}\left(U\right)\right]A_c^{(1)}.\label{eq.A2_2}
\end{align}

The expressions for $P_c^{(1)}$ and $P_c^{(2)}$ are more complicated to write in terms of the matrices $\Lambda$ and $U$. In general terms, considering that the six vertices of the hexagon have positions $\VEC{r}_{i}$, we obtain: 
\begin{align}
P_c^{(1)} = & \sum_{i=1}^6 {P_{c_i}^{(1)}}, \label{eq.P1_11}\\
{P_{c_i}^{(1)}} = & \sqrt{\alpha_i^2 + \beta_i^2},\label{eq.P1_12}\\
P_c^{(2)} = & {P_{c}^{(1)}} + \epsilon M_c^{(1)} + \epsilon^2 M_c^{(2)},\label{eq.P2_1}
\end{align}
with
\begin{align} 
\alpha_i & = \lambda_{xx}{x_{i+1,i}}^{(0)} + \lambda_{xy}{y_{i+1,i}}^{(0)} \label{Alphai},\\
\beta_i & =  \lambda_{yx}{x_{i+1,i}}^{(0)} + \lambda_{yy}{y_{i+1,i}}^{(0)},\label{Betai}
\end{align}
where we use $\VEC{r}_{i+1,1} = \VEC{r}_{i+1} - \VEC{r}_{i}$, assuming the vertices ordered clockwise. The terms $M_c^{(1)}$ and $M_c^{(2)}$ are given by
\begin{align} 
M_c^{(1)} & = \sum_{i=1}^6 \frac{1}{P_{c_i}^{(1)}} \left[\alpha_i^2 u_{xx} + \beta_i^2 u_{yy}+\alpha_i \beta_i \left(u_{xy} + u_{yx}\right) \right], \\
M_c^{(2)}  &= \sum_{i=1}^6 \frac{1}{P_{c_i}^{(1)}}  \left[  u_{xx}^2 \left( \frac{\alpha_i^2}{2}-\frac{\alpha_i^4}{2P_{c_i}^{(1)^2}}\right) + u_{yy}^2\left(\frac{\beta_i^2}{2}-\frac{\beta_i^4}{2P_{c_i}^{(1)^2}} \right) \right. \nonumber \\
&  +u_{xy}^2 \left(\frac{\beta_i^2}{2}-\frac{\alpha_i^2 \beta_i^2}{2P_{c_i}^{(1)^2}} \right) + u_{yx}^2\left(\frac{\alpha_i^2}{2}-\frac{\alpha_i^2 \beta_i^2}{2P_{c_i}^{(1)^2}} \right) \nonumber\\
& + u_{xx}u_{xy}\left(\alpha_i \beta_i - \frac{\alpha_i^3 \beta_i}{P_{c_i}^{(1)^2}}\right) + u_{yy}u_{yx}\left(\alpha_i\beta_i - \frac{\alpha_i \beta_i^3}{P_{c_i}^{(1)^2}} \right)\nonumber \\
& + u_{xx}u_{yx}\left(-\frac{\alpha_i^3 \beta_i}{P_{c_i}^{(1)^2}} \right) + u_{yy}u_{xy}\left(-\frac{\alpha_i^3 \beta_i}{P_{c_i}^{(1)^2}} \right)\nonumber \\
& \left. + u_{xx}u_{yy}\left(-\frac{\alpha_i^2 \beta_i^2}{P_{c_i}^{(1)^2}}\right) +u_{xy}u_{yx}\left(-\frac{\alpha_i^2 \beta_i^2}{P_{c_i}^{(1)^2}}\right) \right].
\end{align}

Now, following a similar procedure as in the previous section we can compute all the energy contributions.
The contribution proportional to $K_A$ is 
 \begin{align}
E_A = &\sum_c \frac{K_A}{2}\left( A_c^{(2)} - A_{0c} \right)^2, \nonumber \\
=&\sum_c \frac{K_A}{2}A_{0c}^2\left[   \left(1+\epsilon \mathrm{tr}\left(U\right) + \epsilon^2 \mathrm{det}\left(U\right)\right)  \mathrm{det}\left(\Lambda \right) - 1        \right]^2,\nonumber \\
=&\sum_c \frac{K_A}{2}A_{0c}^2\left[  \mathrm{det}\left(\Lambda \right) -1+\left(\epsilon \mathrm{tr}\left(U\right) + \epsilon^2 \mathrm{det}\left(U\right)\right)  \mathrm{det}\left(\Lambda \right)        \right]^2,
\label{eq.EA1}
\end{align}
where we obtain that the zeroth, first, and second order terms of $E_A$ are given by
 \begin{align}
 E_{A}^{(0)}  = &\sum_c \frac{K_A }{2}A_{0c}^2 \left(\mathrm{det}\left(\Lambda\right) -1\right)^2 , \label{AreaEnergy01}\\
E_{A}^{(1)}  = &\sum_c K_A A_{0c}^2 \mathrm{det}\left(\Lambda\right) \left[\mathrm{det}\left(\Lambda\right) - 1\right]\mathrm{tr}\left(U\right) , \label{AreaEnergy12}\\
E_{A}^{(2)}  = &\sum_c \frac{K_A}{2}A_{0c}^2\left[ \mathrm{det}\left(\Lambda\right)^2 \mathrm{tr}\left(U\right)^2+2\mathrm{det}\left(\Lambda\right)\left(\mathrm{det}\left(\Lambda\right) - 1\right)\mathrm{det}\left(U\right)\right]. \label{AreaEnergy22}
\end{align}
 
Similarly, for the term proportional to $K_P$,
\begin{align}
E_P= &\sum_c \frac{K_P}{2}\left( P_c^{(2)} - P_{0c} \right)^2,\nonumber \\
=& \sum_c \frac{K_P}{2}\left( P_{c}^{(1)} - P_{0c} +  \epsilon M_c^{(1)} + \epsilon^2 M_c^{(2)}
 \right) ^2,
\label{eq.EP2}
\end{align}
and the  zeroth, first, and second order terms of $E_P$ are given by
\begin{align}
E_{P}^{(0)}  =& \sum_c \frac{K_P}{2}\left(P_c^{(1)}-P_{0c}\right)^2, \label{PerimeterEnergy0}\\
E_{P}^{(1)}  =& \sum_c K_P \left(P_c^{(1)}-P_{0c}\right)M_c^{(1)} , \label{PerimeterEnergy1}\\
E_{P}^{(2)}  = &\sum_c \frac{K_P}{2} \left[ 2\left(P_c^{(1)}-P_{0c}\right)M_c^{(2)}+ {M_c^{(1)}}^2 \right]. \label{PerimeterEnergy2}
\end{align}

Finally,  the zeroth, first, and second order terms of $E_J$ are 
\begin{align} 
E_{J}^{(0)}  = &\sum_c   \frac{J}{2} P_{c}^{(1)},&
E_{J}^{(1)}  = &\sum_c   \frac{J}{2} M_{c}^{(1)} , &
E_{J}^{(2)}  = &\sum_c \frac{J}{2} M_{c}^{(2)}.  \label{JEnergy22}
\end{align}

\section{Equations of motion} \label{app.eqmotion}

With periodic boundary conditions, Eq.~(1) from the main text can be written as
\begin{align}
E = \sum_c\frac{K_A}{2} \left(A_{c}-A_{0c} \right)^2 + \sum_c\frac{K_P}{2} \left(P_{c}-P_{0c} \right)^2 + \sum_c \frac{J}{2}P_{c}.
\end{align}

The equations of motion for the vertex are obtained using Eq.~(2) of the main text, which can be written as \begin{align}
\frac{d\VEC{r}_i}{dt} & = \left.\frac{d\VEC{r}_i}{dt}\right|_{A} + \left.\frac{d\VEC{r}_i}{dt}\right|_{P}  + \left.\frac{d\VEC{r}_i}{dt}\right|_{J}. 
\label{eq.eq_mot}
\end{align}

Assuming a polygon of $N$ vertices, we calculate its area using the triangularization method with respect to the vertex $v_1$,
\begin{align}
A_c = & -\sum_{j=2}^{{N-1}} \frac{1}{2} \HAT z \cdot \left(\VEC{r}_{j,1}\times \VEC{r}_{j+1,1}\right), \nonumber \\
=&  -\sum_{j=2}^{{N-1}} \frac{1}{2} \HAT z \cdot \left[\left(\VEC{r}_{j}-\VEC{r}_{1}\right)\times \left(\VEC{r}_{j+1}-\VEC{r}_{1}\right)\right],\nonumber \\
= &   \sum_{j=2}^{{N-1}} \frac{1}{2} \HAT z \cdot \left[-\VEC{r}_{j} \times \VEC{r}_{j+1} +\VEC{r}_{v_1}\times\left( \VEC{r}_{j+1}-\VEC{r}_{j}\right)\right],\nonumber \\
= & \sum_{j=2}^{{N-1}} \frac{1}{2} \HAT z \cdot \left[-\VEC{r}_{j} \times \VEC{r}_{j+1}\right] +   \frac{1}{2} \HAT z \cdot \left[\VEC{r}_{1} \times \left(\VEC{r}_{N} - \VEC{r}_{2} \right) \right], 
\end{align}
where we used that the tissue is in the $x$-$y$ plane, with the vertices in each cell ordered clockwise, and we defined $\VEC{r}_{i,j} = \VEC{r}_i - \VEC{r}_j$ and $r_{i,j} = \VEC{r}_{i,j}/ \vert \VEC{r}_{i,j} \vert$. To compute the energy gradients, it is convenient to write this expression using any vertex to make the triangularization
\begin{align}
A_c = & \sum_{j=2}^{{N-1}} \frac{1}{2} \HAT z \cdot \left[-\VEC{r}_{j} \times \VEC{r}_{j+1}\right] +   \frac{1}{2} \HAT z \cdot \left[\VEC{r}_{i} \times \left(\VEC{r}_{i-1} - \VEC{r}_{i+1} \right) \right], 
\end{align}
where cyclic vertex numbering is used (i.e. $N+1\equiv1$ and $-1\equiv N$). Then,
\begin{align}
\VEC{\nabla}_iA_c = & \frac{1}{2}\VEC{\nabla}_i \left( \HAT z \cdot \left[\VEC{r}_{i}\times\left( \VEC{r}_{i_c-1}-\VEC{r}_{i+1}\right)\right] \right), \nonumber\\
=& \frac{1}{2}\VEC{\nabla}_i \left[ x_i \left(y_{i-1}-y_{i+1}\right)-y_i \left( x_{i-1}-x_{i+1} \right)  \right],\nonumber\\
 = &  \frac{1}{2} \left(y_{i-1}-y_{i_c+1}\right)\HAT x - \frac{1}{2}\left(x_{i-1}-x_{i+1}\right)\HAT y = \frac{1}{2} \VEC{r}_{i+1,i-1}\times \HAT z.
\end{align}

Also,  the perimeter and its gradient with respect to the position of the vertex $i$ of the same polygon are given by
\begin{align}
P_c = & \sum_{j=1}^{N} \vert \VEC{r}_{j+1,j}\vert,\\
\VEC{\nabla}_i P_c = & \VEC{\nabla}_i \left(\vert \VEC{r}_{i+1,i}\vert+\vert \VEC{r}_{i-1,i}\vert \right)
= -\frac{\VEC{r}_{i+1,i}}{r_{i+1,i}} - \frac{ \VEC{r}_{i_c-1,i}}{r_{i-1,i}}.
\end{align}

Finally, the different terms of Eq.~\eqref{eq.eq_mot} are
\begin{align}
\frac{d\VEC{r}_i}{dt}^{(A)}& = -\sum_c K_A \left(A_{c}-A_{0c} \right)\VEC{\nabla}_iA_c, \nonumber \\
& = -\sum_c K_A \left(A_{c}-A_{0c} \right)\frac{1}{2}\left\{\VEC{r}_{i_c+1,i_c-1}\times \HAT z \right\}, \nonumber \\
&=  \sum_c K_A \left(A_{c}-A_{c0} \right)\frac{1}{2}\left\{ \VEC{r}_{i_c-1,i_c+1}\times \HAT z\right\},\label{eq.vA}\\
\frac{d\VEC{r}_i}{dt}^{(P)} & = - \sum_c K_P \left(P_{c}-P_{c0} \right)\VEC{\nabla}_iP_c, \nonumber \\
& = \sum_c K_P \left(P_{c}-P_{c0} \right)\left(\frac{{\VEC{r}_{i_c+1,i}}}{r_{i_c+1,i}} + \frac{\VEC{r}_{i_c-1,i}}{r_{i_c-1,i}} \right),\label{eq.vP}\\
\frac{d\VEC{r}_i}{dt}^{(J)} & = - \sum_c \frac{J}{2} \VEC{\nabla}_iP_c,\nonumber \\
& = \sum_c \frac{J}{2} \left(\frac{\VEC{r}_{i_c+1,i}}{r_{i_c+1,i}} + \frac{\VEC{r}_{i_c-1,i}}{r_{i_c-1,i}} \right)\label{eq.vJ},
\end{align}
where Eqs.~\eqref{eq.vA}, \eqref{eq.vP}, and \eqref{eq.vJ} consider a sum over the three cells at which the vertex $i$ belongs to, and $i_c+1$ and $i_c-1$ refer to the next and previous vertex to $i$, in clockwise counting, belonging to cell $c$.

\section{Short and long time scales} \label{app.timescales}

By performing a simple dimensional analysis we can obtain the relevant time scales of the dynamics, and define useful \textit{short time} and  \textit{long time} values,  $\tau_s$ and  $\tau_l$, respectively. The first one allows us to detect the beginning of the instability, while the second allows the non-linear terms, which saturate the eventual instabilities, to act.

We analyze the energy of a single hexagonal cell of equilibrium side $a_0$. At time $t=0$ it is deformed isotropically such that the new side is $a = a_0 +  a_1$, with $a_1\ll a_0$. The area (equilibrium area) and perimeter (equilibrium perimeter) are $3\sqrt{3}a^2/2$ ($3\sqrt{3}a_0^2/2$) and $6a$ ($6a_0$), respectively. To simplify, we consider $J=0$, in which case the energy of the cell is
\begin{align} 
E = \frac{K_A}{2} \frac{27}{4} \left(2a_0 a_1  + a_1^2 \right)^2 + \frac{K_P}{2} \left(6 a_1 \right)^2.
\end{align}
According to the dynamics of the vertex model, the cell side evolves as 
\begin{align} 
\dot{a_1}\sim -\derpar{E}{a_1} & =- \left[\frac{27}{4}K_A \left(2a_0 a_1  + a_1^2 \right)\left(2a_0 +2a_1\right) + 36K_Pa_1 \right],\nonumber\\
& =  -\left[\left(\frac{27}{\tau_A} + \frac{36}{\tau_P}\right)a_1 + \frac{(81/2)}{\tau_A}\frac{ a_1^2}{a_0}  + \frac{(27/2)}{\tau_A}\frac{ a_1^3}{a_0^2} \right],
\end{align}
where we defined $\tau_A = 1/(K_A a_0^2)$ and $\tau_P = 1/(K_P)$.
With the selection of units such that $K_A=a_0=1$, we have that  $\tau_A = 1$ and $\tau_P = 1/p$, which is of order 1. Hence,
\begin{align} 
\dot{a_1} = - \frac{a_1}{1/\left(27 + 36\right)} - \frac{a_1^2}{2/81} - \frac{a_1^3}{2/27}.
\end{align}
Obviously, for a confluent tissue, the linear and non-linear terms change, and there are parameters for which the coefficients change sign and tissue is stable. Nevertheless, the present analysis allows us to extract  the relaxation time scales. The shortest gives the linear evolution, $\tau_1\approx 0.016$, and the other two describe the non-linear terms $\tau_{2}\approx0.025$ and $\tau_{3}\approx0.074$. 
If we consider the short time $\tau_s = 0.025$, the unstable modes will have grown exponentially, allowing us to identify their effect in the form a change in ellipticity. For the long time $\tau_l=0.5$, the non-linear terms have played a role and the system could have reach a steady state if the non-linear terms saturate the instability.

\section{Comparison between regular and irregular tissues} \label{app.comparison}
\begin{figure}[H] 
 \includegraphics[width=1\linewidth]{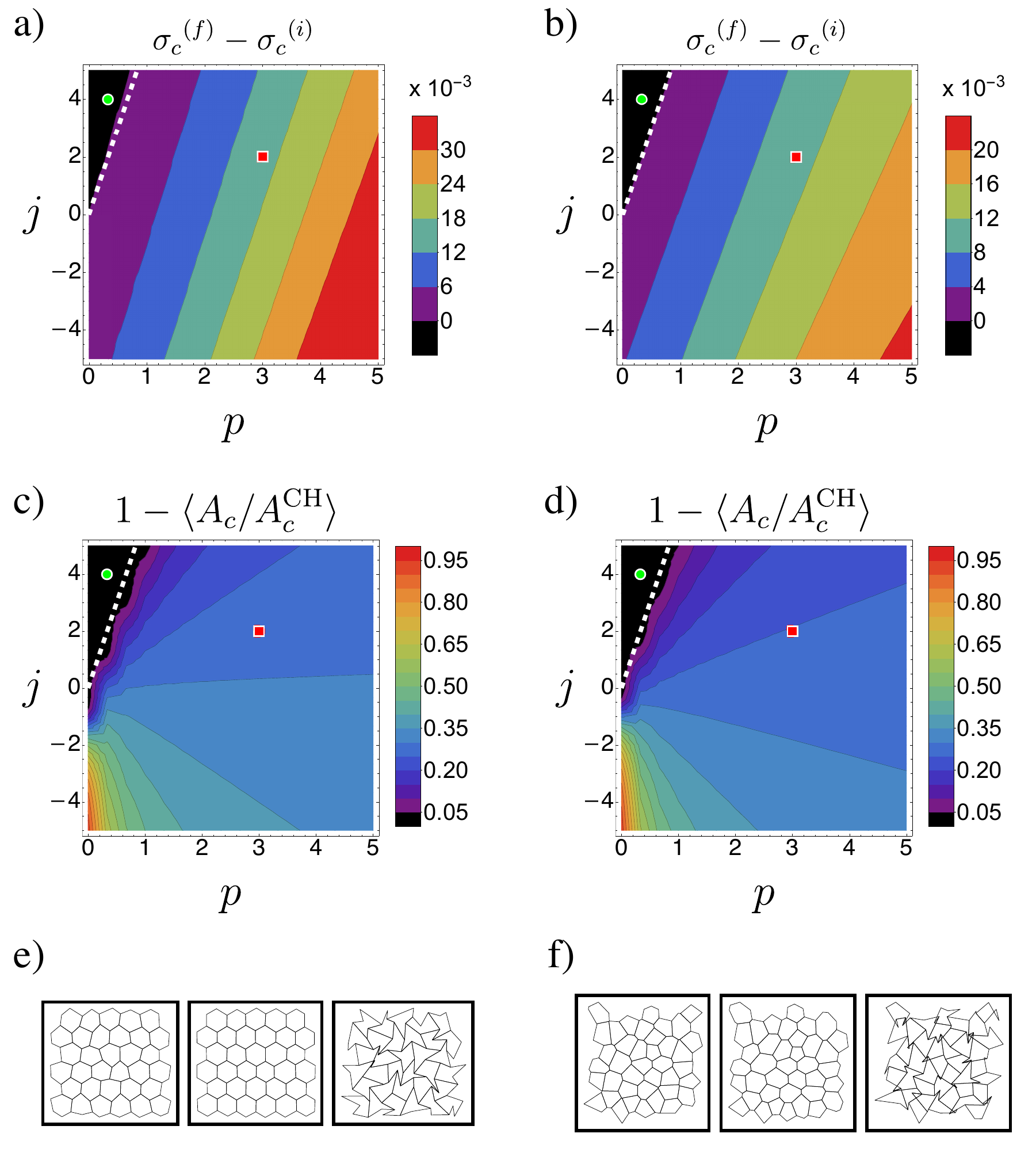}
  \begin{picture}(0,0)(0,0)
\put(25,55){I}
\put(58,55){II}
\put(91,55){III}
\put(152,55){I}
\put(184,55){II}
\put(218,55){III}
 \end{picture}\vspace{-0.2cm}
 \caption{ Tissue instabilities obtained in simulations of $N=3000$ hexagonal cells with $10\%$ of Gaussian noise over the regular positions (left) and irregular cells (right) (three different tissues considered), under cell target perimeter activity, with $\lambda_P=+1/2$ and $\lambda_A=0$. Same 
representation as in Fig.~\ref{fig.resultados_coupledmodes}.}
 \label{fig.simP0_1}
 \end{figure}

\begin{figure}[H] 
    \includegraphics[width=1\linewidth]{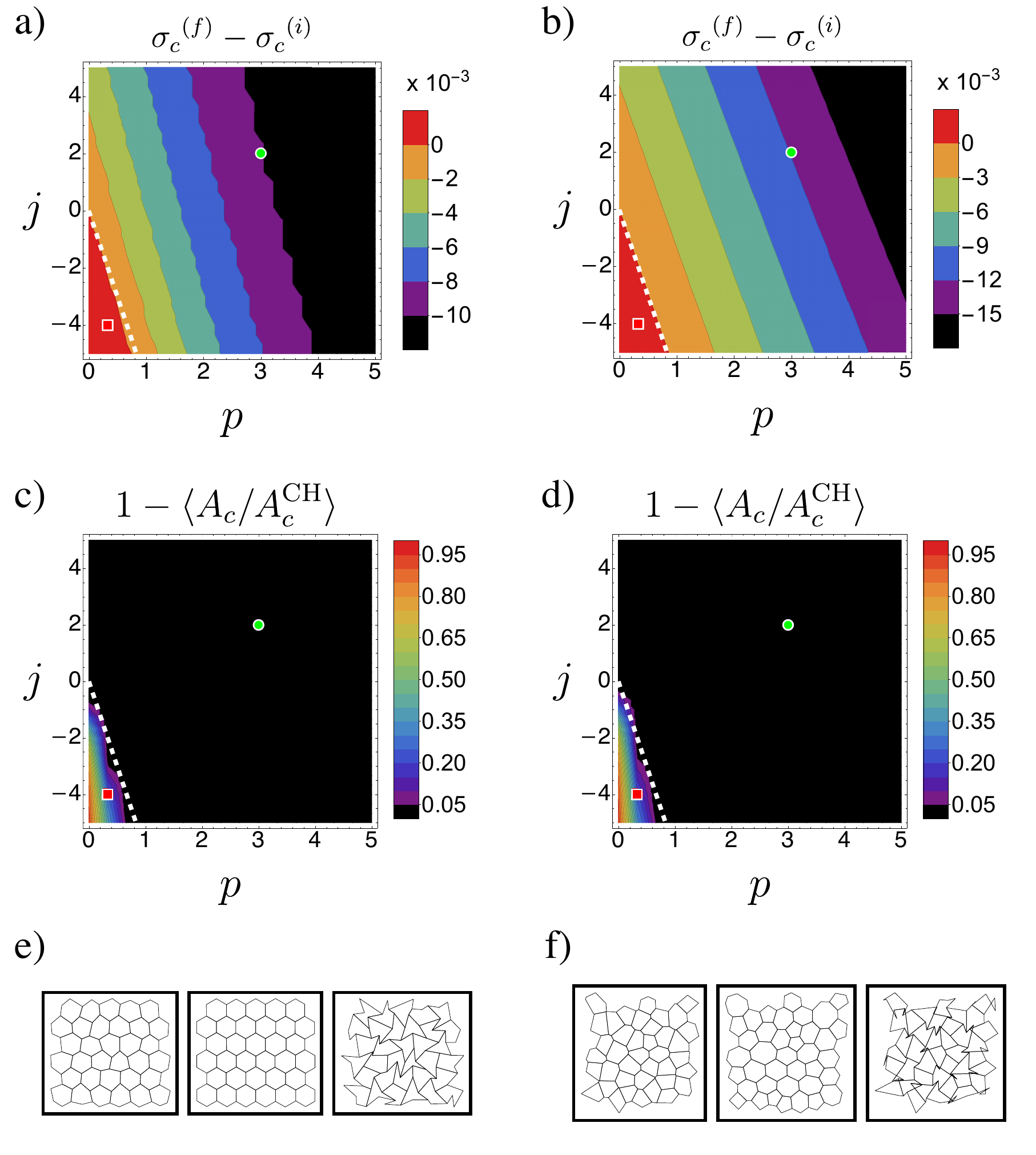} 
     \begin{picture}(0,0)(0,0)
\put(25,55){I}
\put(58,55){II}
\put(91,55){III}
\put(151,55){I}
\put(184,55){II}
\put(218,55){III}
 \end{picture}\vspace{-0.2cm}
 \caption{  Tissue instabilities obtained in simulations of $N=3000$ hexagonal cells with $10\%$ of Gaussian noise over the regular positions (left) and irregular cells (right) (three different tissues considered), under cell target perimeter activity, with $\lambda_P=-1/2$ and $\lambda_A=0$. Same 
representation as in Fig.~\ref{fig.resultados_coupledmodes}.}
 \label{fig.simP0_2}
 \end{figure}

To compare the dynamics of regular and irregular tissues, we performed simulations for both cases. The results for target perimeter activity, with $\lambda_P=+1/2$ (predicted line: $j=6p$) and $\lambda_P=-1/2$ (predicted line: $j=-6p$), can be seen in Figs.~\ref{fig.simP0_1} and \ref{fig.simP0_2}, respectively. Although the detailed geometry of the cells change, the flattening parameter and the measure of non-convexity  agree remarkable well between regular and irregular tissues, showing that the long wavelength approximation is valid. From Figs.~\ref{fig.simP0_1}b and \ref{fig.simP0_2}b it is seen that $\lambda_P=-1/2$ achieves lower values for the standard deviation of the flattening parameter, which results in more rounded cells [Fig.~\ref{fig.simP0_2}f(II) versus Fig.~\ref{fig.simP0_1}f(II)].

\onecolumngrid
 %\begin{widetext}

\section{Examples of non-diagonal $\mu$-matrices} \label{app.nondiagonal}

Using the expressions in Appendix~\ref{ap.energy} it is possible to derive the $\mu$-matrix for different cases. Here, we present some examples where the resulting matrix is non-diagonal, needing the analysis described in Section \ref{sec.analysisnondiag} to determine the unstable modes.

For an anisotropic deformation, characterized by a   $60\%$ horizontal contraction and  $40\%$ vertical expansion, $\Lambda =  \begin{pmatrix}
0.4 & 0\\
0 & 1.4\end{pmatrix}$. The $\mu$-matrix is 
\begin{equation}
\mu_{60/40} = \begin{pmatrix}
0.246+0.019j+1.090p & 0 & 0 & 1.632p\\
0 & 0.246+0.193j-0.110p & -0.143+0.081p & 0 \\
0 & -0.143j+0.081p & -0.246+0.212j-0.121p & 0\\
1.632p & 0 & 0 & 0.381+2.420p
\end{pmatrix}.
\end{equation}
The transition line is given by $j =0.569p$. 
Simulation results for irregular tissues can be seen in Fig.~\ref{fig.resultados2}.
 
For a tissue under a pure deviatoric deformation, $\Lambda =  \begin{pmatrix}
0.5 & 0\\
0 & 1.5
\end{pmatrix}$,  the $\mu$-matrix is  
\begin{equation}
\mu_\text{dev} = \begin{pmatrix}
0.188+0.027j+1.150p & 0 & 0 & 1.824p\\
0 & 0.188+0.206j+0.120p & -0.145j-0.085p & 0\\
0 & -0.145j-0.085p & -0.188+0.233j+0.136p & 0\\
1.824p & 0 & 0 &0.938 + 2.932p
\end{pmatrix}.
\end{equation}
The associated matrix $e_\text{ps}$ is obtained [Eqs.~\eqref{eq.e11}, \eqref{eq.e12}, and \eqref{eq.e22}]  and we compute the curve in parameter space where the minimum eigenvalue of $e_\text{ps}$ changes its sign. Equivalently we search when the determinant vanishes, finding the linear relation $j = -0.583p$. Note that, although the $\Lambda$ and $\mu$ matrices are similar to the previous case, the transition line is radically different. Simulation results for irregular tissues can be seen in Fig.~\ref{fig.act_otros}.

Finally, for a tissue subject to  a pure shear pre-deformation, $\Lambda =  \begin{pmatrix}
1 & 0.5\\
0.5 & 1
\end{pmatrix}$,  the $\mu$-matrix is 
 \begin{equation}
\mu_\text{ps} = \begin{pmatrix}
0.19+0.16j+0.13p & 0.03j + 0.15p & 0.16j + 0.12p & 0.18p\\
0.03j +0.15p & 0.19+0.08j+1.48p & -0.01j-0.01p & 2.07p\\
0.16j + 0.12p & -0.01j-0.01p & -0.19 + 0.24j + 0.18p & 0\\
0.18p & 2.07p & 0 & 0.94 + 3.02p
\end{pmatrix}.
 \end{equation}
 The line at which the minimum eigenvalue of $e_\text{dev}$ changes its sign is given by $j = -0.769p$. Simulation results for irregular tissues can be seen in Fig.~\ref{fig.act_otros}.

 %\end{widetext}

\twocolumngrid 

%\bibliography{biblio-FP2}

\end{document}